# Impedance Network of Interconnected Power Electronics Systems: Impedance Operator and Stability Criterion

Chen Zhang, Marta Molinas, *Member, IEEE*, Atle Rygg, Xu Cai

*Abstract*—Impedance is an intuitive and efficient way for dynamic representation of power electronics devices. One of the evident strengths, when compared to other small-signal methods, is the natural association with circuit theory. This makes them possible to be connected through basic circuit laws. However, careful attention should be paid when making this association since the impedances obtained through linearization are local variables, often referred to locally defined reference frames. To allow the operations of these impedances using basic circuit laws, a unified/global reference has to be defined. Though this issue was properly addressed on the state-space models, a thorough analysis and a clarification regarding the unified impedances and stability effects are still missing. This paper aims to bridge this gap by introducing the Impedance Operator (IO) and associated properties to the development of impedance networks. First, the IO for both the AC coupled and AC/DC coupled systems are presented and verified through impedance measurements in PSCAD/EMTDC. Then, three types of impedance network-based stability criterions are presented along with a clarification on the consistency of stability conclusions. Finally, the Nyquist-based analysis is explored, regarding the sensitivity to partition points, to open the discussion on the identification of systems' weak points.

I. INTRODUCTION

Nowadays, power electronics devices, e.g. the voltage source converters (VSCs), have been widely adopted for the grid-integration of renewable energies [1] as well as the interconnection of asynchronous AC grids by means of the high-voltage-dc (HVDC) technology [2]. In addition to the bulk power system, power electronics devices in micro-grids [3] also exhibit superior capability in increasing overall efficiency and flexibility. Hence, the power electronics devices are widespread in modern power systems, giving rise to a significant concern on the new dynamics and stability issues. Among them, the small signal stability issue, e.g. in a manner of wide-band oscillation [4], is the most stringent one since it has been experienced frequently in the field, e.g. the wind parks [5] and solar power plants [6].

Recently, numerous studies and efforts have been realized in this area. One can broadly classify them into two groups: the state-space model with eigenvalue analysis (e.g. [4], [7] and [8]) and the impedance-based model with frequency domain analysis (e.g. [10]-[17]). Of which, the state-space method is well-established to some extent since it has been utilized to study the electromechanical oscillations of traditional power systems for a long time. On the other hand, the impedance-based method has become more prevalent in recent years due to its convenience in derivation and interpretation. Typically, e.g. the impedances of grid-connected devices can be obtained through either analytical modeling or field measurements, moreover, they are a kind of "impedance" to a certain extent, hence new dynamics can be interpreted and more easily understood in view of circuit analyses.

Currently, the impedance modeling and stability analysis of a single grid-tied VSC is extensively discussed. There are various techniques to derive the VSC impedances from different viewpoints, a thorough review is presented in [18]. The most representative modeling methods are the *dq* impedance modeling (e.g. [10], [11] and [12]) and the sequence impedance modeling (e.g. [16] and [17]), which are derived respectively from the linearized systems in *dq* domain and sequence domain. For symmetrical three-phase systems, they are generally two-by-two matrices with nonzero off-diagonal elements. This implies that the impedances of actively controlled VSCs (referred to as the "active impedance") are coupled multi-input and multi-output (MIMO) systems, and this coupling of a typical VSC can be interpreted as the mirror frequency coupling (MFC) effect [13] or equivalently the sequence coupling effect [17]. It is worth to notice that this coupling effect turns out to be important for stability analysis and should not be overlooked, particularly for the low-frequency dynamics analysis. Once the VSC and the grid impedance are derived, the stability condition caused by the interaction between VSC and the grid can be evaluated via the (Generalized) Nyquist criterion [19]. For this analysis, the source and load subsystems partitioned at a specific point (typically the point of common coupling, PCC) should be defined first, and then certain conditions on the poles of the source and load have to be considered [20]. Thereupon, the stability of the grid-VSC system can be concluded by inspecting the eigen-loci and counting the encirclements of the critical point (-1, 0 j).

Once the impedances of individual components (e.g. the VSC and the transmission lines) are derived, it is easy to associate and establish the impedance network to perform the multi-converter analysis, or, generally speaking, the interconnected systems analysis. For example, [5] and [21] have analyzed the sub/super synchronous oscillation of wind farms via impedance manipulation, whereas in [22] the harmonic resonance issue is focused. However, this intuitive association of the currently developed impedances (e.g. [10] and [13]) with circuit analyses is not accurate since the impedances obtained through linearization are only representing the local behaviors of devices, i.e. the currents/voltages characterizing the impedances are local variables. Therefore, those impedances cannot be directly connected for which a unified/global reference frame has to be defined. Though this issue has been addressed before in the state-space modeling (e.g. [8]) and some relevant remarks in [21], a thorough analysis, clarification and validation in terms of impedance network is missing and of crucial importance for stability studies of large networks with large number of power electronics units.

Therefore, this work aims to bridge this gap by introducing the Impedance Operator (IO) and associated properties to the formation of impedance networks. Of which, the IO is defined in this paper as the procedures to perform mathematical operation of locally evaluated active impedances before formulating the impedance network with the basic circuit rules (e.g. series, parallel). The rest of the paper is organized as follows:

In section II, the problem of impedance operator is put forward by introducing the characteristics of active impedances. Then, the IO for both AC coupled and AC/DC coupled power electronics systems are established, whereby its property and impact on circuit manipulations are discussed. Based on the developed IO, the impedance networks of the AC and AC/DC coupled systems can be established with the knowledge of basic circuit laws. Once this has been done, stability can be evaluated in different ways, e.g. the Nyquist-based or the circuit analysis-based approach. Therefore, a discussion on the impedance network-based stability criterions is provided in section III, their consistencies with respect to stability conclusions are clarified. To further address the importance of IO on stability analysis, section IV presents some case-studies, along with a discussion on the finding of system's vulnerable points. Finally, section V draws the main conclusions.

## II. IMPEDANCE OPERATOR FOR INTERCONNECTED POWER ELECTRONICS SYSTEMS

### A. *Properties of the active and passive impedances*

Fig. 1 shows a typical grid-tied VSC system, the VSC control system is usually comprised of three parts: the inner current-control-loop (CCL), the phase-locked loop (PLL) and the outer control loop, of which the CCL and the PLL are fundamental controls for a grid-synchronized VSC, whereas the outer loop can be the dc voltage control or active and reactive power control according to the operating mode.

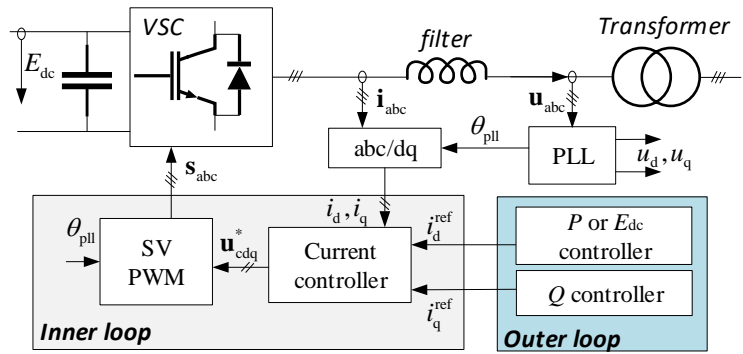

Fig. 1 Schematic of a typical grid-connected VSC system

Recently, extensive efforts have been dedicated to the PLL in particular in the context of weak AC grids. Its effects on either VSC's small signal [23] or large signal [24] stability is discussed in depth. Among them, the most evident effect of PLL in view of impedances is the *dq* asymmetry property [25]. For example, a VSC impedance in *dq* domain is generally represented as [12]:

$$\begin{bmatrix} U_d(s) \\ U_q(s) \end{bmatrix} = \underbrace{\begin{bmatrix} Z_{dd}(s) & Z_{dq}(s) \\ Z_{qd}(s) & Z_{qq}(s) \end{bmatrix}}_{\mathbf{Z}_{dq}(s)} \begin{bmatrix} I_d(s) \\ I_q(s) \end{bmatrix} \quad (1)$$

thereby the definition of *dq* symmetry is the condition: $Z_{dd}(s) = Z_{qq}(s)$ and $Z_{dq}(s) = -Z_{qd}(s)$. The definition for *dq* asymmetry is the opposite, i.e. the condition is not met. According to the definition, intuitively, most of the passive impedances are *dq* symmetric, e.g. the inductance: $Z_L(s) = \begin{bmatrix} sL & -\omega_1 L \\ \omega_1 L & sL \end{bmatrix}$. However, the active impedances are mostly *dq* asymmetric since the majority of them intrinsically have the PLL effects.

An evident consequence of the *dq* asymmetry is that it can introduce frequency couplings to the system, e.g. if a VSC is perturbed by a small sequence component at $\omega_p$ from three-phase, then the response will not only present a component at $\omega_p$ but also a component at $-\omega_p + 2\omega_1$. This is essentially the MFC effect as mentioned before, and better illustration is achieved if the modified sequence domain (MSD) impedance [13] is adopted, which is invented from the *dq* impedance through the symmetrical decomposition [26], e.g.:

$$\mathbf{Z}_{pn}(s) = \mathbf{T}_{sym} \cdot \mathbf{Z}_{dq}(s) \mathbf{T}_{sym}^{-1} = \begin{bmatrix} Z_{pp}(s) & Z_{pn}(s) \\ Z_{np}(s) & Z_{nn}(s) \end{bmatrix} \quad (2)$$

where

$$\begin{cases} Z_{pp}(s) = \dfrac{Z_{dd}(s) + Z_{qq}(s)}{2} + \dfrac{Z_{qd}(s) - Z_{dq}(s)}{2} j \\ Z_{pn}(s) = \dfrac{Z_{dd}(s) - Z_{qq}(s)}{2} + \dfrac{Z_{qd}(s) + Z_{dq}(s)}{2} j \\ Z_{np}(s) = \left(Z_{pn}(-s)\right)^* \\ Z_{nn}(s) = \left(Z_{pp}(-s)\right)^* \end{cases} \quad (3)$$

and $\mathbf{T}_{sym} = \dfrac{1}{2}\begin{bmatrix} 1 & j \\ 1 & -j \end{bmatrix}$. It is noted that the lower case notation "pn" is adopted to distinguish the modified sequence domain from the original sequence domain. According to (2), it is clear that $Z_{pn}(s) = Z_{np}(s) = 0$ is obtained if the system is *dq* symmetric, indicating there is no MFC effect. In this way, the *dq* symmetric property is intuitively associated with the MFC effect through the MSD impedances. Due to this benefit, all the impedances in the later discussion are developed in MSD, moreover, this will not lose the generality of analysis since the MSD and the *dq* impedance are linearly dependent.

It is noted that the MFC effect is a general and distinctive characteristic of active impedances compared to the passive ones, thus special attention has to be paid when manipulating them. For which, in the following, the IO and associated properties are introduced and discussed for both of the AC coupled and AC/DC coupled systems.

### B. Impedance operator for AC coupled systems

In the first place, taking the system in Fig. 2 (a) as an example, without loss of generality, both of the VSCs are *PQ* controlled with the adoption of the PLL (see appendix for the models), whereas the dc voltage is assumed constant. The power lines are linear passive elements, in this study, they are considered to be inductance dominant.

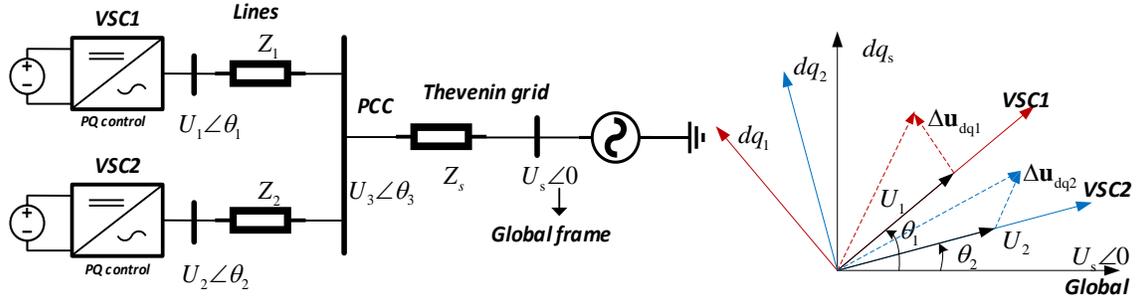

(a) Schematic of an AC coupled system    (b) Illustration of vectors and reference frames

Fig. 2 A simple interconnected AC power electronics system for the study

Since the VSC impedances are only representing the local behaviors of the currents and voltages (see $\Delta \mathbf{u}_{dq1}, \Delta \mathbf{u}_{dq2}$ in Fig. 2 (b)), they cannot be connected directly, (e.g. $\Delta \mathbf{u}_{dq}^{series} \neq \Delta \mathbf{u}_{dq1} + \Delta \mathbf{u}_{dq2}$, if the series-connection of them is considered). In order to manipulate them as traditional circuit elements, the currents/voltages characterizing the impedances should be consistent with each other, which means they should refer to a unified/common reference frame, e.g. the $dq_s$ reference frame of the infinite-bus bar (in fact the common reference frame can be chosen arbitrarily). This is fulfilled by reference frame transformation.

According to Fig. 2 (b), $\Delta \mathbf{u}_{dq1}$ of VSC1 with respect to $dq_1$ can be transformed into the common reference by:

$$t-domain : \begin{bmatrix} \Delta u_{d\_dq1} \\ \Delta u_{q\_dq1} \end{bmatrix} = \begin{bmatrix} \cos\theta_1 & \sin\theta_1 \\ -\sin\theta_1 & \cos\theta_1 \end{bmatrix} \begin{bmatrix} \Delta u_{d\_dqs} \\ \Delta u_{q\_dqs} \end{bmatrix} \quad (4)$$

where $\theta_1$ is the load angle of VSC1 relative to the infinite bus-bar $U_s\angle 0$. Since the transformation is linear-time-invariant, the relationships in s-domain, as well as the MSD are derived:

$$s-domain: \begin{bmatrix} U_{d\_dq1}(s) \\ U_{q\_dq1}(s) \end{bmatrix} = \underbrace{\begin{bmatrix} \cos\theta_1 & \sin\theta_1 \\ -\sin\theta_1 & \cos\theta_1 \end{bmatrix}}_{\mathbf{T}_{dq}(\theta_1)} \begin{bmatrix} U_{d\_dqs}(s) \\ U_{q\_dqs}(s) \end{bmatrix}$$

$$MS-domain: \begin{bmatrix} U_{p\_dq1}(s+j\omega_1) \\ U_{n\_dq1}(s-j\omega_1) \end{bmatrix} = \underbrace{\begin{bmatrix} e^{-j\theta_1} & 0 \\ 0 & e^{j\theta_1} \end{bmatrix}}_{\mathbf{T}_{rot}(\theta_1)} \begin{bmatrix} U_{p\_dqs}(s+j\omega_1) \\ U_{n\_dqs}(s-j\omega_1) \end{bmatrix} \quad (5)$$

Based on this transformation, the local impedance can be generally transformed to the global one as:

$$MSD: \mathbf{Z}_{pn}^{global}(s) = \mathbf{T}_{rot}(-\theta_1)\mathbf{Z}_{pn}^{local}(s)\mathbf{T}_{rot}(\theta_1)$$
$$dq: \mathbf{Z}_{dq}^{global}(s) = \mathbf{T}_{dq}(-\theta_1)\mathbf{Z}_{dq}^{local}(s)\mathbf{T}_{dq}(\theta_1) \quad (6)$$

$\mathbf{Z}_{pn}^{local}(s)$ and $\mathbf{Z}_{dq}^{local}(s)$ are reintroduced from (2). In essence, the process for deriving the global impedance according to (6) denotes the IO of this work, apparently lacking such process will result in inaccurate impedances. In the later analysis, only the MSD impedances are focused.

Further, several properties of the IO on the MSD impedances are revealed:

**P.1** the passive impedances (strictly speaking, the *dq* symmetric impedances) are invariant in terms of IO;

**P.2** the IO only affect the off-diagonals of the active impedances (strictly speaking, the *dq* asymmetric impedances) by shifting their phases;

**P.3** the eigen-loci of the active/passive impedances are not affected by the IO.

They can be easily proven by expanding (6) explicitly as:

$$\mathbf{Z}_{pn}^{global}(s) = \begin{bmatrix} Z_{pp}^{local}(s) & Z_{pn}^{local}(s)e^{j2\theta_1} \\ Z_{np}^{local}(s)e^{-j2\theta_1} & Z_{nn}^{local}(s) \end{bmatrix} \quad (7)$$

from (7), P.1 is obtained since condition holds: $Z_{pn}^{local}(s) = Z_{np}^{local}(s) = 0$ due to the *dq* symmetry. P.2 is derived since the IO only positively and negatively rotate the off-diagonals by double of the load angles. P.3 can be justified according to the definition of the eigen-loci:

$$\det(\lambda\mathbf{I} - \mathbf{Z}_{pn}^{global}) = \det(\lambda\mathbf{I} - \mathbf{T}_{rot}(-\theta_1)\mathbf{Z}_{pn}^{local}(s)\mathbf{T}_{rot}(\theta_1))$$
$$= \det\mathbf{T}_{rot}(-\theta_1)\cdot\det(\lambda\mathbf{I} - \mathbf{Z}_{pn}^{local})\cdot\det\mathbf{T}_{rot}(\theta_1) \quad (8)$$
$$= \det(\lambda\mathbf{I} - \mathbf{Z}_{pn}^{local}) = 0$$

According to the above-mentioned process, the impedances seen from the PCC (see Fig. 2 (a)) with (denoted by $\mathbf{Z}_{pn\_pcc}^{global}$) and without IO (denoted by $\mathbf{Z}_{pn\_pcc}^{local}$) IO are derived:

$$\mathbf{Z}_{pn\_pcc}^{global} = \left(\mathbf{Z}_1 + \mathbf{Z}_{pn\_vsc1}^{global}\right) \| \left(\mathbf{Z}_2 + \mathbf{Z}_{pn\_vsc2}^{global}\right)$$
$$\mathbf{Z}_{pn\_pcc}^{local} = \left(\mathbf{Z}_1 + \mathbf{Z}_{pn\_vsc1}^{local}\right) \| \left(\mathbf{Z}_2 + \mathbf{Z}_{pn\_vsc2}^{local}\right) \quad (9)$$

where $\mathbf{Z}_{pn\_vsc1}^{global} = \mathbf{T}_{rot}(-\theta_1)\mathbf{Z}_{pn\_vsc1}^{local}(s)\mathbf{T}_{rot}(\theta_1)$ and $\mathbf{Z}_{pn\_vsc2}^{global} = \mathbf{T}_{rot}(-\theta_2)\mathbf{Z}_{pn\_vsc2}^{local}(s)\mathbf{T}_{rot}(\theta_2)$. Clearly, $\mathbf{Z}_{pn\_pcc}^{global} \neq \mathbf{Z}_{pn\_pcc}^{local}$ if the VSCs are loaded (i.e. $\theta_1, \theta_2 \neq 0$).

To further illustrate the IO effects on impedance characteristics, the aggregated PCC impedances developed with and without IO (i.e. $\mathbf{Z}_{pn\_pcc}^{global}$ and $\mathbf{Z}_{pn\_pcc}^{local}$) are compared with impedance measurements from simulations in PSCAD/EMTDC. Control parameters of the VSC1 and the VSC2 are the same. The operating points are different, where $P_{vsc1} = 1.0\ p.u.$ and $P_{vsc2} = -0.5\ p.u.$, the results are shown in Fig. 3.

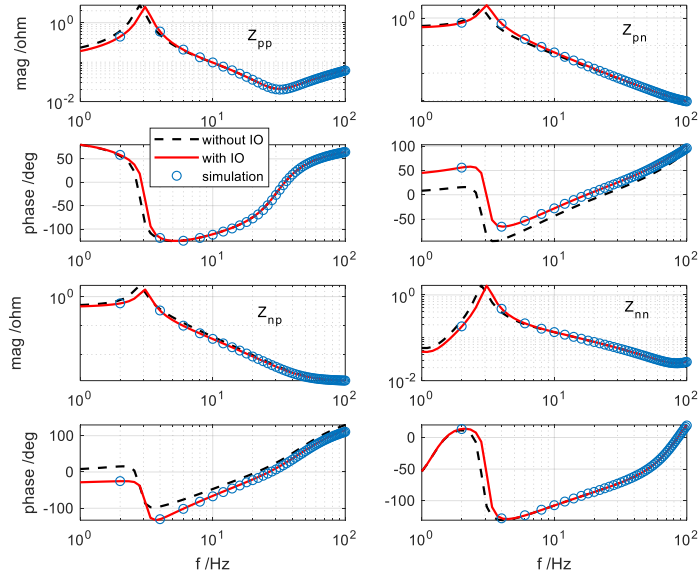

Fig. 3 Impedance comparisons of the AC coupled system ($P_{vsc1}$ = 1.0 p.u.; $P_{vsc2}$ = -0.5p.u., PLL = 20 Hz, CC = 300 Hz, PQ = 20 Hz, $Z_s$ = 0.25 j p.u., $Z_1 = Z_2 = 0.1$j p.u. frequency-sweeping is from 2Hz to 100 Hz with an increment of 2 Hz)

It is identified that the aggregated PCC impedance (i.e. $\mathbf{Z}_{pn\_pcc}^{global}$) with the introduced IO is consistent with simulations, whereas the one without IO (i.e. $\mathbf{Z}_{pn\_pcc}^{local}$) presents some discrepancies on amplitudes and phases. Further, it is noted that, though the IO only affects the individual component's phases of off-diagonals (according to P1), its effects on the aggregated impedance are widespread, see the magnitude and phase responses of the diagonals are also affected by the IO. This is because of the series and parallel operations, more importantly, the different load angles of

VSC1 and VSC2. Since all the elements of the aggregated impedances are affected by the IO, the eigen-loci, as well as the accuracy in stability analysis, are expected to be affected as well, this will be discussed in section IV. B.

*C. Impedance operator for AC/DC coupled systems*

In addition to the AC transmission lines, the HVDC transmission system is an alternative for the grid-integration of VSCs, e.g. the off-shore wind farms. A representative but simplified study system is presented in Fig. 4 (a). The voltage angles e.g. $\theta_1, \theta_2, \theta_3, \theta_{hvdc1}$ are referred to an arbitrary reference fame. e.g. $\angle\theta=0$, which is also the common reference frame in this case. In the following, the IO and associated properties for such AC/DC coupled system will be developed.

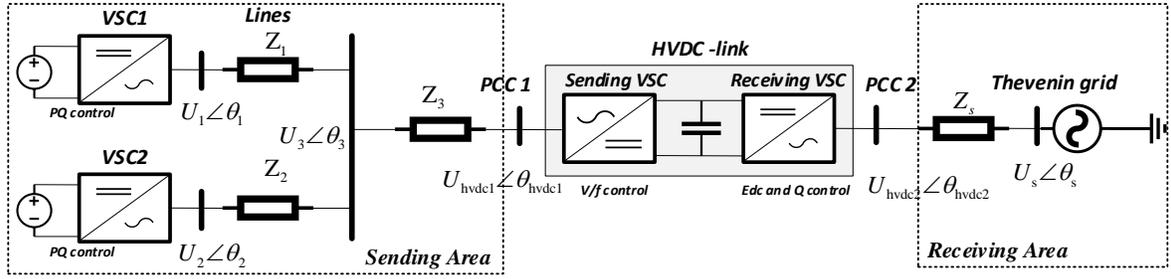

(a) Schematic of the AC/DC coupled system

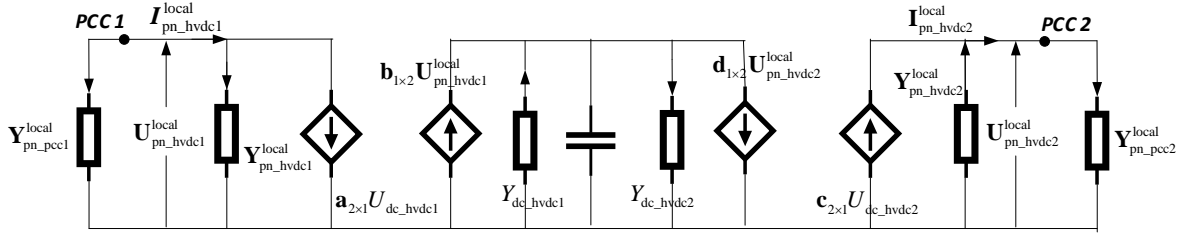

(b) The equivalent circuit of the AC/DC coupled system

Fig. 4 A simple interconnected AC/DC power electronics system for the study

Typically, the sending-end VSC adopts V/f control, of which the voltage at PCC1 is controlled and a constant frequency is applied. Whereas the receiving-end VSC adopts the dc voltage control with the capability of reactive power compensations. For better analysis, both of them can be compactly represented by three-port modules [27], e.g. for the sending-end VSC (AC currents flow into the VSC is positive, dc current flows into the dc-link is positive), it can be represented by (see appendix for the models):

$$\begin{bmatrix} I^{local}_{p\_hvdc1}(s) \\ I^{local}_{n\_hvdc1}(s) \\ I_{dc\_hvdc1}(s) \end{bmatrix} = \begin{bmatrix} \mathbf{Y}^{local}_{pn\_hvdc1}(s) & \mathbf{a}_{2\times1}(s) \\ \mathbf{b}_{1\times2}(s) & Y_{dc\_hvdc1}(s) \end{bmatrix} \begin{bmatrix} U^{local}_{p\_hvdc1}(s) \\ U^{local}_{n\_hvdc1}(s) \\ U_{dc\_hvdc1}(s) \end{bmatrix} \quad (10)$$

where $\mathbf{I}_{pn\_hvdc1}^{local}$, $\mathbf{U}_{pn\_hvdc1}^{local}$ are local variables with respect to the local reference frame provided by $\angle\theta_{hvdc1}$.

Since the dc-side variables are irrelevant to reference frames, the AC/DC IO for the HVDC system (i.e. without loss of generality, the AC/DC coupled system) can be developed by modifying the AC IO as:

$$\mathbf{T}_{rot\_hvdc}(\theta_{hvdc1}) = \begin{bmatrix} \mathbf{T}_{rot}(\theta_{hvdc1}) & \mathbf{0}_{2\times 1} \\ \mathbf{0}_{1\times 2} & 1 \end{bmatrix} \quad (11)$$

Applying this AC/DC IO on (10) yields:

$$\begin{bmatrix} I_{p\_hvdc1}^{global} \\ I_{n\_hvdc1}^{global} \\ I_{dc\_hvdc1} \end{bmatrix} = \mathbf{T}_{rot\_hvdc}(-\theta_{hvdc1}) \begin{bmatrix} \mathbf{Y}_{pn\_hvdc1}^{local} & \mathbf{a}_{2\times 1} \\ \mathbf{b}_{1\times 2} & Y_{dc\_hvdc1} \end{bmatrix} \mathbf{T}_{rot\_hvdc}(\theta_{hvdc1}) \begin{bmatrix} U_{p\_hvdc1}^{global} \\ U_{n\_hvdc1}^{global} \\ U_{dc\_hvdc1} \end{bmatrix}$$

$$= \begin{bmatrix} \mathbf{T}_{rot}(-\theta_{hvdc1})\mathbf{Y}_{pn\_hvdc1}^{local}\mathbf{T}_{rot}(\theta_{hvdc1}) & \mathbf{T}_{rot}^{hvdc}(-\theta_{hvdc1})\mathbf{a}_{2\times 1} \\ \mathbf{b}_{1\times 2}\mathbf{T}_{rot}(\theta_{hvdc1}) & Y_{dc\_hvdc1} \end{bmatrix} \begin{bmatrix} U_{p\_hvdc1}^{global} \\ U_{n\_hvdc1}^{global} \\ U_{dc\_hvdc1} \end{bmatrix} \quad (12)$$

clearly, this IO does not affect the dc component $Y_{dc\_hvdc1}$ but all the AC/DC coupled elements are affected.

Further, this AC/DC IO can be simplified if the whole system is analyzed at the dc side of the HVDC-link. Assuming the sending area (see Fig. 4 (a)) impedance seen from PCC 1 in the global reference frame (i.e. $\angle 0$) is $\mathbf{Y}_{pn\_pcc1}^{global}$, if written explicitly, it can be developed from (9) as:

$$\mathbf{Y}_{pn\_pcc1}^{global} = \left[\mathbf{Z}_1 + \mathbf{T}_{rot}(-\theta_1)\mathbf{Z}_{pn\_vsc1}\mathbf{T}_{rot}(\theta_1)\right]^{-1} + \left[\mathbf{Z}_2 + \mathbf{T}_{rot}(-\theta_2)\mathbf{Z}_{pn\_vsc2}\mathbf{T}_{rot}(\theta_2)\right]^{-1} \quad (13)$$

Then, substituting it into (12) and eliminating the ac nodes yields:

$$I_{dc\_hvdc1} = \left[\mathbf{b}_{1\times 2}\left(\mathbf{Y}_{pn\_pcc1}^{local} - \mathbf{Y}_{pn\_hvdc1}^{local}\right)^{-1}\mathbf{a}_{2\times 1} + Y_{dc\_hvdc1}\right]\cdot U_{dc\_hvdc1}$$

$$= -Y_{dc\_send}\cdot U_{dc\_hvdc1} \quad (14)$$

where $\mathbf{Y}_{pn\_pcc1}^{local} = \mathbf{T}_{rot}(\theta_{hvdc1})\mathbf{Y}_{pn\_pcc1}^{global}\mathbf{T}_{rot}(-\theta_{hvdc1})$ is with respect to the local reference frame provided by the $\angle\theta_{hvdc1}$. It is also noted that, the $\mathbf{b}_{1\times 2}$ and $\mathbf{a}_{2\times 1}$ are no longer affected by the AC/DC IO as (12).

Thus, for the dc side analysis of the HVDC-link, only the ac impedances of the sending area are affected. As a result, the AC IO for the $i$ th ac side impedance is $\mathbf{T}_{rot}(\theta_i - \theta_{hvdc1})$, where $\theta_i$ is the voltage angle with respect to the common reference frame (i.e. $\angle\theta = 0$). This is fulfilled by expanding $\mathbf{Y}_{pn\_pcc1}^{local}$ using (13), where $\mathbf{Z}_{pn\_vsc1}^{local} = \mathbf{T}_{rot}(\theta_{hvdc1} - \theta_1)\mathbf{Z}_{pn\_vsc1}\mathbf{T}_{rot}(\theta_1 - \theta_{hvdc1})$ and $\mathbf{Z}_{pn\_vsc2}^{local} = \mathbf{T}_{rot}(\theta_{hvdc1} - \theta_2)\mathbf{Z}_{pn\_vsc2}\mathbf{T}_{rot}(\theta_2 - \theta_{hvdc1})$ are obtained.

For the receiving area analysis, the common reference frame can be chosen as the voltage angle of the Thevenin grid, i.e. $\angle \theta_s = 0$ (see Fig. 4 (a)). Likewise, the three-port module of the receiving-end VSC can be established as (ac currents flow out of VSC is positive, dc current flows into the converter is positive, see appendix for the models):

$$\begin{bmatrix} I_{\text{p\_hvdc2}}^{\text{local}} \\ I_{\text{n\_hvdc2}}^{\text{local}} \\ I_{\text{dc\_hvdc2}} \end{bmatrix} = \begin{bmatrix} \mathbf{Y}_{\text{pn\_hvdc2}}^{\text{local}} & \mathbf{c}_{2\times 1} \\ \mathbf{d}_{1\times 2} & Y_{\text{dc\_hvdc2}} \end{bmatrix} \begin{bmatrix} U_{\text{p\_hvdc2}}^{\text{local}} \\ U_{\text{n\_hvdc2}}^{\text{local}} \\ U_{\text{dc\_hvdc2}} \end{bmatrix} \quad (15)$$

By introducing the receiving area (see Fig. 4 (a)) impedance seen from PCC 2 in the common reference frame, i.e. $\mathbf{Y}_{\text{pn\_pcc2}}^{\text{global}}$, the dc-side impedance of the receiving-end VSC is obtained as:

$$\begin{aligned} I_{\text{dc\_hvdc2}} &= \left[ \mathbf{d}_{1\times 2} \left( \mathbf{Y}_{\text{pn\_pcc2}}^{\text{local}} - \mathbf{Y}_{\text{pn\_hvdc2}}^{\text{local}} \right)^{-1} \mathbf{c}_{2\times 1} + Y_{\text{dc\_hvdc2}} \right] \cdot U_{\text{dc\_hvdc2}} \\ &= Y_{\text{dc\_rec}} \cdot U_{\text{dc\_hvdc2}} \end{aligned} \quad (16)$$

where $\mathbf{Y}_{\text{pn\_pcc2}}^{\text{local}} = \mathbf{T}_{\text{rot}}(\theta_{\text{hvdc1}}) \mathbf{Y}_{\text{pn\_pcc2}}^{\text{global}} \mathbf{T}_{\text{rot}}(-\theta_{\text{hvdc2}})$ is the admittance with respect to the local reference frame provided by $\angle \theta_{\text{hvdc2}}$. Similar to results of the sending-area, the AC/DC IO only affects the ac side impedances, of which, the IO for the $j$ th ac side impedance is: $\mathbf{T}_{\text{rot}}(\theta_j - \theta_{\text{hvdc2}})$, $\theta_j$ is voltage angle in the receiving area.

This IO is also valid if the HVDC link is analyzed at the ac side, e.g. the sending-end VSC is:

$$\begin{bmatrix} I_{\text{p\_hvdc1}}^{\text{local}}(s) \\ I_{\text{n\_hvdc1}}^{\text{local}}(s) \end{bmatrix} = \left( \mathbf{Y}_{\text{pn\_hvdc1}}^{\text{local}}(s) + \frac{\mathbf{a}_{2\times 1}(s)\mathbf{b}_{1\times 2}(s)}{Y_{\text{dc\_rec}}(s) + sC_{\text{cap}} - Y_{\text{dc\_hvdc1}}(s)} \right) \begin{bmatrix} U_{\text{p\_hvdc1}}^{\text{local}}(s) \\ U_{\text{n\_hvdc1}}^{\text{local}}(s) \end{bmatrix} \quad (17)$$

Similarly, the receiving-end VSC is:

$$\begin{bmatrix} I_{\text{p\_hvdc2}}^{\text{local}}(s) \\ I_{\text{n\_hvdc2}}^{\text{local}}(s) \end{bmatrix} = \left( \mathbf{Y}_{\text{pn\_hvdc2}}^{\text{local}}(s) + \frac{\mathbf{c}_{2\times 1}(s)\mathbf{d}_{1\times 2}(s)}{Y_{\text{dc\_send}}(s) + sC_{\text{cap}} - Y_{\text{dc\_hvdc2}}(s)} \right) \begin{bmatrix} U_{\text{p\_hvdc2}}^{\text{local}}(s) \\ U_{\text{n\_hvdc2}}^{\text{local}}(s) \end{bmatrix} \quad (18)$$

It is noted that, for the ac side analysis of the sending-end VSC, the dc side impedance of the receiving-end VSC (i.e. $Y_{\text{dc\_rec}}(s)$) is developed according to its own reference frame, in which the ac side impedances within the receiving area have to be manipulated according to the introduced IO: $\mathbf{T}_{\text{rot}}(\theta_j - \theta_{\text{hvdc2}})$. Similarly, if the analysis is conducted on the ac side of the receiving-end VSC, all the ac impedances within the sending area have to be manipulated according to the IO: $\mathbf{T}_{\text{rot}}(\theta_i - \theta_{\text{hvdc1}})$.

In summary, the HVDC-link decouples the ac systems of the two sides in terms of reference frames. For each ac system, the local ac impedances should refer to the reference frame provided by the corresponding end of the VSC-

HVDC, e.g. $\angle\theta_{hvdc1}$ for the sending area. Once this has been done, all the ac and dc side impedances are unified and can be manipulated with basic circuit laws in either dc side or ac side, thereby the equivalent circuit of such AC/DC coupled system can be drawn in Fig. 4 (b).

To further illustrate the IO effects, the dc-side impedances of the AC/DC coupled system, developed with and without the introduced IO, are compared together with the impedance measurements. The results are shown in Fig. 5.

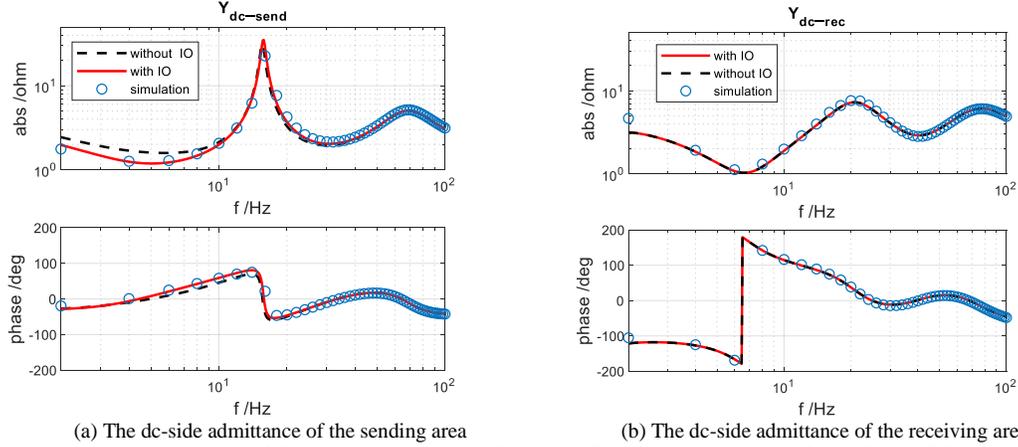

(a) The dc-side admittance of the sending area    (b) The dc-side admittance of the receiving area

Fig. 5 Impedance comparisons of the AC/DC coupled system (for VSC1 and VSC2: PQ control = 10 Hz, PLL = 20 Hz, CC = 300 Hz, P = 1.0. p.u.; for VSC-HVDC, dc voltage control = 50 Hz, Q control = 10 Hz, PLL = 10 Hz, ac grid SCR = 4; frequency sweeping is from 2Hz to 100 Hz with an increment of 2 Hz; $Z_1 = Z_2 = Z_3 = 0.1$ j p.u.)

Clearly, the dc side impedances developed with the IO are matched with simulations. In detail, for the sending area, the dc side impedance with absent of the IO leads to evident errors in amplitudes and phases, particularly in the low frequency range (e.g. below 10 Hz in Fig. 5 (a) ). This implies that they will predict different stability margin or even draw the wrong stability conclusion if the system has potential resonances in this frequency range. On the other hand, Fig. 5 (b) presents that the dc side impedances of the receiving area that developed with IO and without IO are consistent. However, this consistency is not a general conclusion since in this scenario the receiving area only contains a Thevenin equivalent AC grid, whose impedance is passive and invariant in terms of the impedance operators (according to P.1). If there are also actively controlled devices presented in this area, e.g. VSCs, the dc side impedances with and without IO will not be identical, which is similar to case of Fig. 5 (a).

## III. IMPEDANCE NETWORK-BASED STABILITY CRITERIONS

Once the correct IO is applied, the impedance network of an interconnected system can be established either through the basic circuit laws or systematically by the Norton equivalence-based admittance model. Taking the system in Fig. 2 (a) as an example, the small-signal admittance network can be constructed as:

$$\begin{bmatrix} \mathbf{I}_1(s) \\ \mathbf{I}_2(s) \\ \mathbf{I}_s(s) \end{bmatrix} = \underbrace{\begin{bmatrix} \mathbf{Y}_{sys}^{sub1} = \begin{bmatrix} \mathbf{Y}_{pn\_vsc1} + \mathbf{Y}_1 & \mathbf{0}_{2\times2} \\ \mathbf{0}_{2\times2} & \mathbf{Y}_{pn\_vsc2} + \mathbf{Y}_2 \end{bmatrix} & \mathbf{Y}_{sys}^{sub2} = \begin{bmatrix} -\mathbf{Y}_1 \\ -\mathbf{Y}_2 \end{bmatrix} \\ \mathbf{Y}_{sys}^{sub3} = \begin{bmatrix} -\mathbf{Y}_1 & -\mathbf{Y}_2 \end{bmatrix} & \mathbf{Y}_{sys}^{sub4} = (\mathbf{Y}_1 + \mathbf{Y}_2 + \mathbf{Y}_s) \end{bmatrix}}_{\mathbf{Y}_{sys}(s)} \begin{bmatrix} \mathbf{U}_1(s) \\ \mathbf{U}_2(s) \\ \mathbf{U}_3(s) \end{bmatrix} \quad (19)$$

where, $\mathbf{I}_s(s) = \mathbf{Y}_s(s)\mathbf{U}_s(s)$ is the Norton equivalence of the ac grid. $\mathbf{I}_1(s)$ and $\mathbf{I}_2(s)$ are the independent current sources of VSC1 and VSC2, whereas $\mathbf{U}_1(s)$ and $\mathbf{U}_2(s)$ are their terminal voltages. It should be noted that all the elements in $\mathbf{Y}_{sys}(s)$ are already manipulated by the correct IO, where the unified reference frame is provided by $U_s\angle 0$. Also, it is worth mentioning those admittances are in the MSD, which means $\mathbf{Y}_1, \mathbf{Y}_2, \mathbf{Y}_3$ are diagonal matrices due to the *dq* symmetric property. It is seen that the MSD impedances are not only superior in physical interpretations but also making calculation easier compared to the *dq* impedances.

Given by this admittance network model, stability can be evaluated by either the Nyquist criterion or the circuit based approaches. However, each of them has some conditions and restrictions, and if not properly considered, wrong stability conclusions may be drawn, and a clarification on this respect is necessary. Besides, in later analysis, all the equivalences and associated criteria will be universally derived from this admittance matrix.

*A. Nyquist criterion-based approach*

To apply the Nyquist-based stability analysis, a partition point of the source and load subsystem has to be defined, typically the PCC in Fig. 2 (a). If the PCC is chosen as the partition point, then the equivalent circuit of the source and load can be drawn in Fig. 6 (a). In which, the source subsystem is straightforward, i.e. $\mathbf{Z}_{Source}(s) = \mathbf{Z}_s(s)$. The load subsystem can be derived from (19) by replacing the Nothon equivalent grid with an independent current source injection at PCC. This is fulfilled by replacing $\mathbf{I}_s(s)$ with $\mathbf{I}_{inj}(s)$ and removing the source admittance $\mathbf{Y}_s$ from $\mathbf{Y}_{sys}^{sub4}$:

$$\begin{bmatrix} \mathbf{I}_1(s) \\ \mathbf{I}_2(s) \\ \mathbf{I}_{inj}(s) \end{bmatrix} = \begin{bmatrix} \mathbf{Y}_{sys}^{sub1} & \mathbf{Y}_{sys}^{sub2} \\ \mathbf{Y}_{sys}^{sub3} & \mathbf{Y}_{sys}^{sub4} - \mathbf{Y}_s \end{bmatrix} \begin{bmatrix} \mathbf{U}_1(s) \\ \mathbf{U}_2(s) \\ \mathbf{U}_3(s) \end{bmatrix} \quad (20)$$

thereby, the characteristic impedance/admittance of the load subsystem can be calculated by setting $\mathbf{I}_1(s) = \mathbf{I}_2(s) = \mathbf{0}$, and measuring the voltage response $\mathbf{U}_3(s)$:

$$\mathbf{I}_{inj}(s) = \left( \mathbf{Y}_{sys}^{sub4} - \mathbf{Y}_s - \mathbf{Y}_{sys}^{sub3} \left( \mathbf{Y}_{sys}^{sub1} \right)^{-1} \mathbf{Y}_{sys}^{sub2} \right) \cdot \mathbf{U}_3(s) = \mathbf{Y}_{Load}(s) \cdot \mathbf{U}_3(s) \quad (21)$$

Besides, the Norton equivalent current in Fig. 6 (a) can also be obtained from (20) by short-circuiting the PCC, i.e. $\mathbf{U}_3(s) = \mathbf{0}$ and measuring $-\mathbf{I}_{inj}(s)$, which is $\mathbf{I}_{Load}(s) = -\mathbf{Y}_{sys}^{sub3}\left(\mathbf{Y}_{sys}^{sub1}\right)^{-1}\begin{bmatrix}\mathbf{I}_1(s)\\ \mathbf{I}_2(s)\end{bmatrix}$.

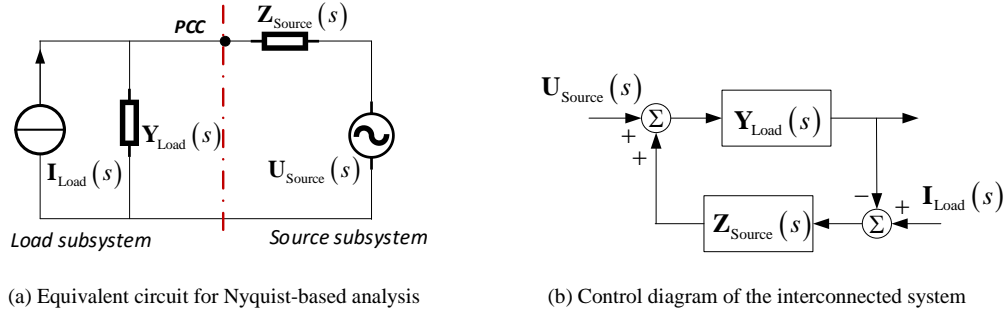

(a) Equivalent circuit for Nyquist-based analysis  (b) Control diagram of the interconnected system

Fig. 6 Nyquist-based analysis of an interconnected system

Since interconnection of the source and load forms a closed-loop system in Fig. 6 (b), stability can be evaluated through the loop-gain according to the Nyquist Criterion, i.e. plotting the eigen-loci of the minor loop gain: $\mathbf{L}_{AC}(s) = \mathbf{Z}_{Source}(s) \cdot \mathbf{Y}_{Load}(s)$ and counting the encirclements of the critical point (-1, 0j), which is rephrased as:

*Stability criterion 1(SC1)* [20]*: if $\mathbf{L}_{AC}(s)$ does not contain any right half plane (RHP) poles, then the closed-loop system is stable if and only if the eigen-loci of $\mathbf{L}_{AC}(s)$ does not encircle the critical point (-1, 0 j).*

However, it is noted that as the partition point moves away from the VSCs' terminals, the load $\mathbf{Y}_{Load}(s)$ may haveright-half-plane (RHP) poles due to the interconnection with other elements (see the further discussion in section IV.A). Thereupon, the system is stable if and only if the number of clockwise encirclements of the (-1,0 j) equals to the number of the RHP poles, otherwise it is unstable. Therefore, it is necessary to check the poles of the subsystems before inspecting the encirclements.

B. *Circuit property-based approach*

As an alternative to the Nyquist criterion, the stability of the closed-loop system can be analyzed through the loop impedance [21] . According to Fig. 6 (a), the loop impedance seen from the voltage source perturbation is:

$$\mathbf{Z}_{Loop}(s) = \mathbf{Z}_{Load}(s) + \mathbf{Z}_{Source}(s) \quad (22)$$

where $\mathbf{Y}_{Load}(s) = \mathbf{Z}_{Load}^{-1}(s)$. Hence, assuming the grid has a small perturbation on the voltage, the current response of the circuit is stable if $\mathbf{Y}_{Load}(s)$ does not have RHP poles. This can be rephrased as:

*Stability Criterion 2 (SC2): The closed-loop system is stable if and only if there are no RHP poles in the $\mathbf{Y}_{Loop}$, or equivalently speaking, there are no RHP zeros in the det($\mathbf{Z}_{Loop}$).*

Instead of calculating the loop impedance, another circuit property-based stability criterion can be obtained directly from the admittance matrix (19). Since the Norton currents are independent inputs, the system is stable if $\mathbf{Z}_{sys}$ does not have RHP poles, which is rephrased as:

*Stability Criterion 3 (SC3): The closed-loop system is stable if and only if there are no RHP poles in the $\mathbf{Z}_{sys}$, or equivalently speaking, there are no RHP zeros in the det($\mathbf{Y}_{sys}$).*

Comparing SC2 and SC3 it can be found that the circuit property based stability criterions are tightly related to the types of circuit equivalences, hence special attention should be paid when applying these criterions.

*C. A comparative analysis of the stability criterions on stability analysis*

In this section, stability criterions, i.e. SC1, SC2 and SC3 are compared regarding their consistency on stability conclusions, for which the system in Fig. 2 (a) is analyzed.

In the first place, a marginally stable case is considered in Fig. 7 (a). It is identified that SC1, SC2 and SC3 are consistent regarding the stability conclusions. In detail, since in this case both the source and load have no RHP poles, the Nyquist plots indicate a marginally stable system. Further, from the zeros-plots of SC2 and SC3 it is observed that, the zeros of det ($\mathbf{Y}_{sys}$) and det ($\mathbf{Z}_{Loop}$) are exactly the same. It is also noticed that this is a pair of left-hand plane (LHP) zeros close to the imaginary axis, implying the same marginally stable condition as the Nyquist plots.

Further, an unstable case is presented in Fig. 7 (b), it is fulfilled by a small increase on the PLL bandwidth of VSC1. Apparently, from the Nyquist plots, it is observed that the system is marginally unstable since the encirclement occurs near the critical point. On the other hand, the zeros-plots of the SC2 and SC3 draw the same stability conclusion due to the presence of a pair of RHP zeros near the imaginary axis.

To verify the results based on analytical models, a time domain simulation is conducted and shown in Fig. 7 (c). It is observed that the system is indeed marginally unstable. Furthermore, the oscillation frequency measured from the simulation is: $f_{measure} = \frac{1}{0.06s} \approx 16.7\ Hz$, which is approximately the same as the one predicted by the unstable zeros: $f_{osc} = \frac{111.7}{2\pi} \approx 17.7\ Hz$, see Fig. 7 (b).

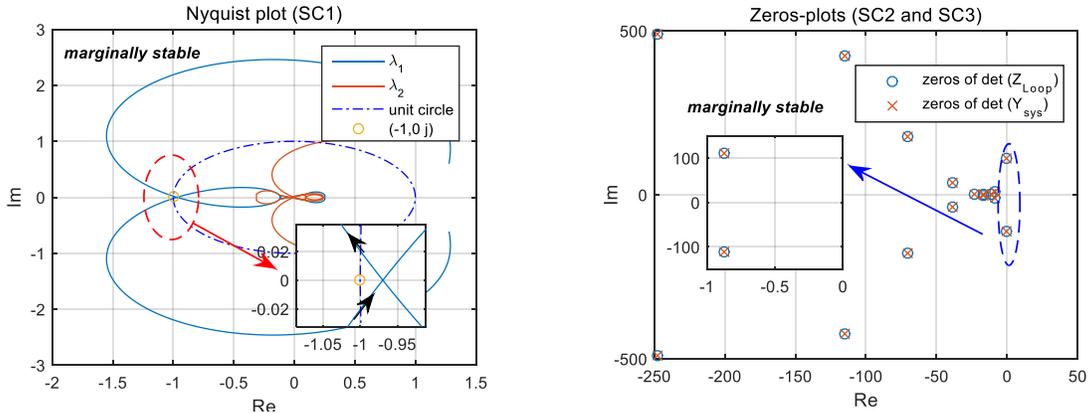

(a) A marginally stable case (VSC1-PLL = 10 Hz, VSC2-PLL = 40 Hz)

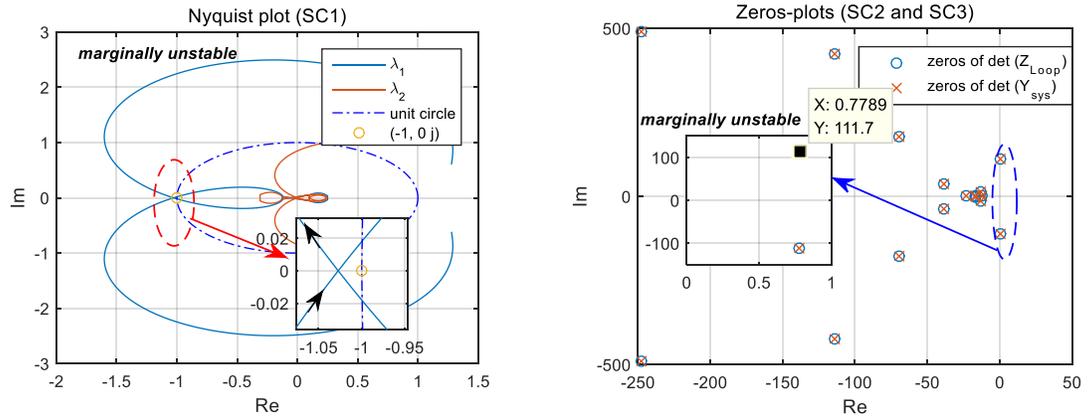

(b) A marginally unstable case (VSC1-PLL = 15 Hz, VSC2-PLL = 40 Hz)

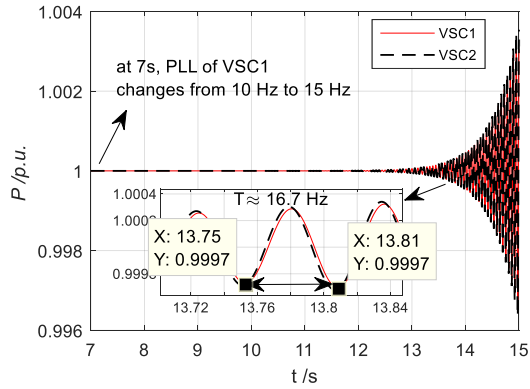

(c) Time domain simulation

Fig. 7 A comparative study of stability criterions (for both VSC1 and VSC2: PQ = 20 Hz, CC = 300 Hz, $P$ = 1.0 p.u.; $Z_1 = Z_2$ = 0.1 j p.u., $Z_s$ = 0.125 p.u.)

Overall, through this comparison analysis, it is confirmed that these stability criterions are effective and consistent on small-signal stability analysis. However, as addressed before, both of them have pros and cons. For the Nyquist-based approach (i.e. SC1), it is illustrative and easy to implement, but the open-loop poles of the loop gain have to be checked before drawing the stability conclusions. For the circuit-based approaches (i.e. SC2 and SC3), though the stability can be concluded straightforwardly through the eigenvalues, the adoption of impedance or admittance poles is tightly related to the types of circuit equivalents.

Although the choice of partition point and the inspection of open-loop poles might be inconveniences for Nyquist-based analysis, it can also provide more information about stability margin than the other two. This feature will be explored further in the next section.

## IV. DISCUSSIONS

### A. Impacts of partition points on open-loop poles

As mentioned before, open-loop poles of the source and the load should be evaluated when applying the Nyquist criterion (i.e. SC1). Taking the equivalent circuit of the source and load system as an example (see Fig. 6 (a), corresponding study system is Fig. 2 (a)), when the partition point is moved from the PCC towards the grid, the equivalent source impedance will be $Z_{Source}^{part}(s) = (1-k_{part})Z_{Source}(s)$, whereas the equivalent load model is: $Y_{Load}^{part}(s) = \left[Y_{Load}^{-1}(s) + k_{part}Z_{Source}(s)\right]^{-1}$, where the partition factor $k_{part}$ is introduced for the measure of the distance from the PCC. Since in this study, the source is the Thevenin equivalent grid, whose impedance is inductive and does not have RHP poles, thus only the open-loop poles of $Y_{Load}^{part}(s)$ is evaluated with varying $k_{part} = 0 \sim 0.9$.

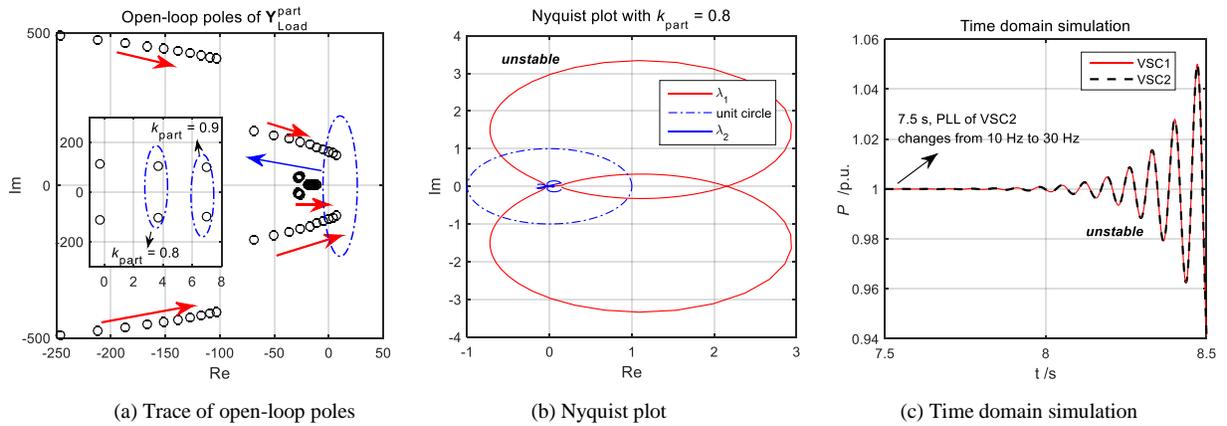

(a) Trace of open-loop poles     (b) Nyquist plot     (c) Time domain simulation

Fig. 8 Impacts of the partition points on the open-loop ploes (for both VSC 1 and VSC2: PQ = 20 Hz, PLL = 30 Hz, CC = 300 Hz, $P$ = 1.0 p.u., $Z_1 = Z_2 = 0.1$ j p.u., $Z_s = 0.1333$ p.u.)

The results are shown in Fig. 8 (a), it is seen that as partition point moves towards the grid (i.e. $k_{part}$ increases), the overall open-loop poles move to the right in the complex plane. Further, when $k_{part} = 0.8$, the RHP open-loop poles are present, which means if the stability is inspected by Nyquist plot, the system is stable if and only if the eigen-loci have one clockwise encirclements of the critical point. However, as illustrated by the Nyquist plots in Fig. 8 (b), there are no encirclements of the critical point, indicating an unstable system.

The time domain simulation is further presented in Fig. 8 (c), where it is clearly shown that the system is unstable after the parameter of the VSC2-PLL is set to the same value as the Nyquist plot. This study addresses the necessity of the check of RHP open-loop poles for Nyquist-based analysis once the new partition point is selected.

### B. Identification of the system's weak point

As mentioned before, the choice of partition point is an additional degree of freedom of Nyquist-based analysis. Since the Nyquist plots can provide more information about stability margin, this feature in combination with the freedom of selecting partition points can be utilized for searching the system's weak point. For example, for finding which VSC in the AC coupled system (see Fig. 2 (a)) is relatively vulnerable with respect to small signal stability.

This analysis is fulfilled by respectively choosing the VSC1's and VSC2's terminal as the partition point for the whole system. As a result, two types of source and load system similar to the Fig. 6 (a) can be developed, based on the Nyquist plots are compared in Fig. 9 (a). It is noted that in this case the source and load systems evaluated at VSC1's and VSC2's terminal have no RHP poles.

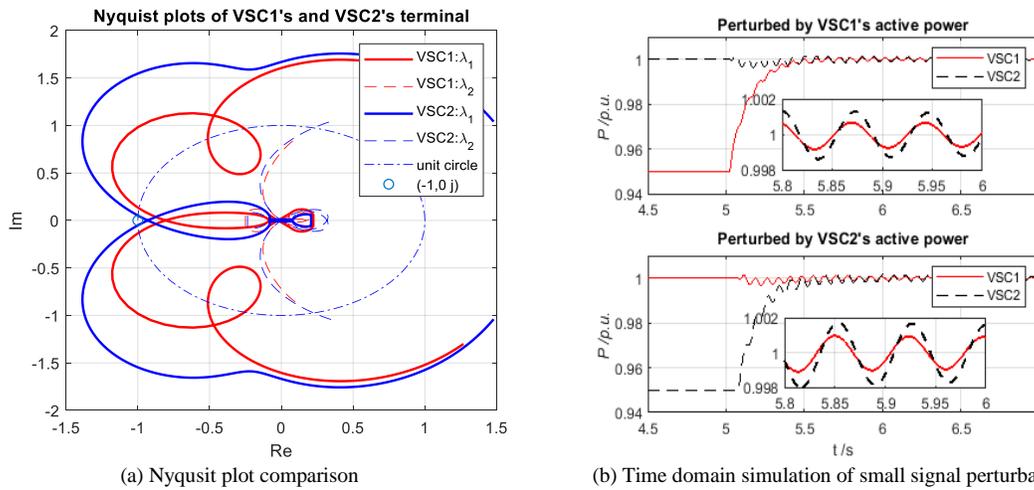

(a) Nyqusit plot comparison  (b) Time domain simulation of small signal perturbations
Fig. 9 Nyquist-based analysis of system's weak point (VSC1: CC = 300 Hz, PLL = 10 Hz, PQ = 10 Hz, $P$ = 1.0 p.u.. VSC2: CC = 240 Hz, PLL = 25 Hz, PQ = 10 Hz, $P$ = 1.0 p.u. Lines: $Z_1 = Z_2 = 0.1$ j p.u., $Z_s = 0.125$ p.u.)

According to Fig. 9 (a), the Nyquist plots (in particular the eigen-loci $\lambda_1$) evaluated at VSC1's terminal exhibit more margin than the VSC2's, which means this partition point is less sensitive to the small signal perturbations compared to the VSC2's. This also implies that the VSC2 is more vulnerable than the VSC1 if the system is perturbed. In Fig. 9 (b), time domain simulations are presented, for which a small step change of either the VSC1's active power

(i.e. from 0.95 pu to 1.0 pu) or the VSC2's is applied for triggering small signal dynamics. It is seen that, regardless of the location of perturbations, the active power response of VSC1 exhibits more damping than the VSC2, proving that VSC2 is indeed less stable.

Once the system's weak point is identified, the impedance-shaping-based methods for small signal stability improvements will become more effective and efficient. For example, in this case, the most straightforward way to improve the overall stability margin is to increase the VSC2' current control bandwidth or reduce its PLL bandwidth.

*C. Impacts of the IO on stability analysis*

In section II, the impacts of the IO are presented in view of frequency responses. To further address its impacts on stability, the AC/DC coupled system in Fig. 4 (a) will be analyzed.

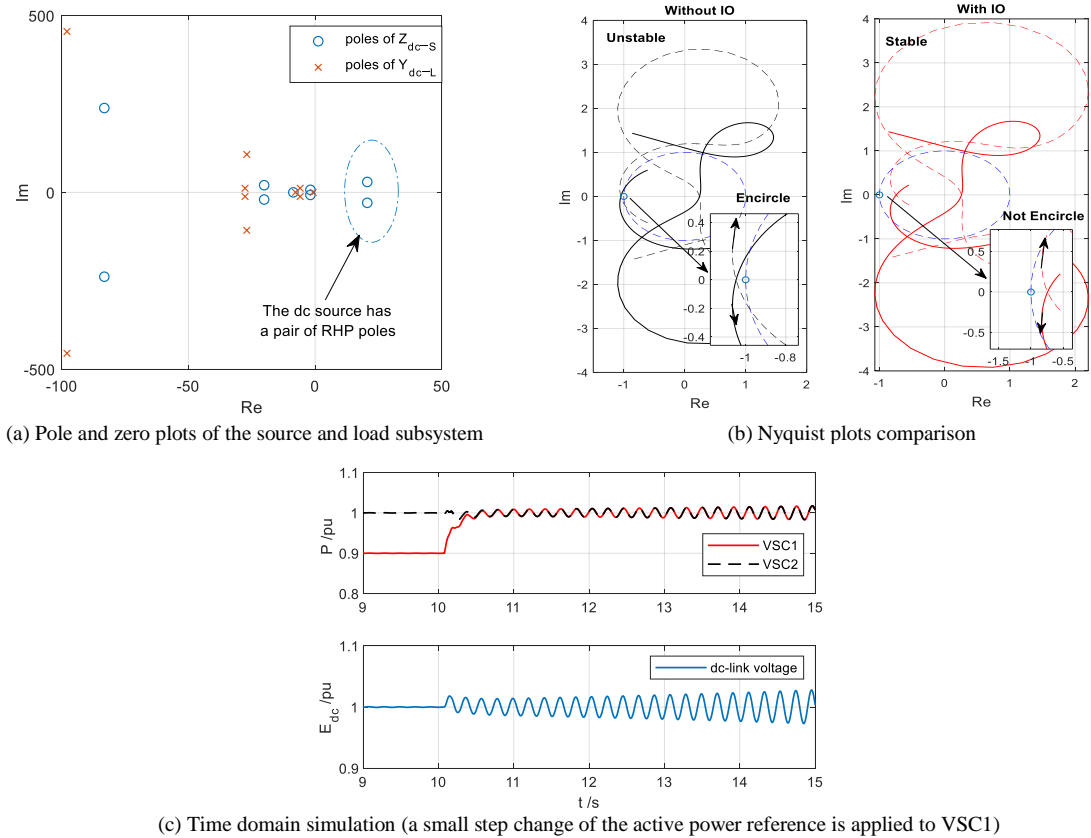

(a) Pole and zero plots of the source and load subsystem  (b) Nyquist plots comparison

(c) Time domain simulation (a small step change of the active power reference is applied to VSC1)

Fig. 10 Nyquist and simulation study of the impedance operator's impacts on stability (VSC1 and VSC2: PLL = 10 Hz, CC = 300 Hz, PQ = 10 Hz, output power = 1 p.u. VSC-HVDC receiving end: PLL = 20 Hz, dc voltage control = 40 Hz, reactive power control = 10 Hz, $Z_1 = Z_2 = 0.1$ j p.u. $Z_3 = 0.15$ j p.u., $Z_s = 0.125$ j p.u.)

Since the dc-side impedances (with IO) seen from the sending and the receiving-VSC have already been developed, i.e. (14) and (16). From which, a source and load system for Nyquist-based analysis (see Fig. 6 (a)) is developed, where the source is defined as $Z_{dc\_S}(s) = Y_{dc\_rec}^{-1}(s)$ and the load is defined as $Y_{dc\_L}(s) = Y_{dc\_send}(s) + s \cdot C_{cap}$.

Similarly, the source and load model without IO can also be obtained, of which the ac side impedances are based on their local reference frames. As a result, their eigen-loci can be plotted and compared.

Before inspecting the Nyquist plots, the open-loop poles of the source and load subsystems are evaluated in Fig. 10 (a), from which it is seen that the source has a pair of RHP poles, whereas the load admittance has no RHP poles. It is worth mentioning that in [28] open-loop poles of the VSC-HVDC system are found when the power flow is inversed. However, in this study, the open-loop poles are presented due to the VSCs control parameters. Due to the presence of a pair of RHP poles, the Nyquist plots have to encircle the critical point (-1, 0 j) once in a counterclockwise manner if the system is stable, otherwise, it is unstable. Based on this criterion and Fig. 10 (b), the eigen-loci with IO draws a stable stability conclusion, whereas the eigen-loci without IO concludes an unstable system due to the fact that there are no encirclements of the critical point.

To check which stability conclusion is correct, time domain simulation of the system is conducted and presented in Fig. 10 (c), in which the small signal dynamics are invoked by a small-step change of the VSC1's active power. Clearly, after a small perturbation, the HVDC-link is unstable as indicated by the dc voltage response. Besides, the active power responses (e.g. after 11 s) of VSC1 and VSC2 are identical. This is because in this case, the configurations of VSC 1 and VSC 2 are exactly the same. In general, this case study addressed the importance of the correct IO for impedance network-based analysis, particularly for the correct stability estimation.

## V. Conclusions

Impedance is an intuitive and effective way for dynamic representation of power electronics devices. However, there are still some concerns of significant importance regarding the formation and stability assessment of the impedance-networks and for which a thorough clarification is crucial in these systems. To address this issue, this paper proposed the IO and associated properties for establishing the impedance networks of both AC coupled and AC/DC coupled systems. The main contributions are:

1) The IO and the resulting impedances for both systems are verified through the impedance measurements in simulations. The importance of the IO on the accuracy of stability analysis is emphasized through a case study, where it is shown that the impedance models without IO can lead to wrong stability conclusions.

2) Once the IO is introduced, impedance networks can be established based on the knowledge of circuit analysis. And the overall stability can be evaluated either through the Nyquist-based method or the circuit property-based methods. In this regard, three types of stability criteria are compared, and their consistency in stability conclusion is justified through case studies. Moreover, the pros and cons for each method are clarified, in particular, the impacts of partition points on the RHP open-loop poles of the Nyquist-based analysis is discussed.

3) Specific to the Nyquist-based analysis, the information of stability margin along with its sensitivity to partition points is further exploited, based on which the system's weak points can be identified and located. This capability of the Nyquist-based analysis for impedance networks could be a promising counterpart to the sensitivity analysis of state-space models.

REFERENCES

[1]. R. Teodorescu, M. Liserre, and P. Rodriguez, "Introduction," in Grid converters for photovoltaic and wind power systems, Chichester, United Kingdom: John Wiley & Sons, 2011, pp. 1–4.
[2]. N. Flourentzou, V. G. Agelidis and G. D. Demetriades, "VSC-Based HVDC Power Transmission Systems: An Overview," in IEEE Transactions on Power Electronics, vol. 24, no. 3, pp. 592-602, March 2009.
[3]. J. M. Guerrero, M. Chandorkar, T. Lee and P. C. Loh, "Advanced Control Architectures for Intelligent Microgrids—Part I: Decentralized and Hierarchical Control," in *IEEE Transactions on Industrial Electronics*, vol. 60, no. 4, pp. 1254-1262, April 2013.
[4]. L. P. Kunjumuhammed, B. C. Pal, C. Oates and K. J. Dyke, "Electrical Oscillations in Wind Farm Systems: Analysis and Insight Based on Detailed Modeling," in IEEE Transactions on Sustainable Energy, vol. 7, no. 1, pp. 51-62, Jan. 2016.
[5]. H. Liu, X.X. Xie, J.B. He, T. Xu, Z. Yu, C. Wang, C.Y. Zhang, "Subsynchronous Interaction Between Direct-Drive PMSG Based Wind Farms and Weak AC Networks," in IEEE Transactions on Power Systems, vol. 32, no. 6, pp. 4708-4720, Nov. 2017.
[6]. C. Li, "Unstable Operation of Photovoltaic Inverter From Field Experiences," in IEEE Transactions on Power Delivery, vol. 33, no. 2, pp. 1013-1015, April. 2018.
[7]. L. Fan, C. Zhu, Z. Miao and M. Hu, "Modal Analysis of a DFIG-Based Wind Farm Interfaced With a Series Compensated Network," in *IEEE Transactions on Energy Conversion*, vol. 26, no. 4, pp. 1010-1020, Dec. 2011.
[8]. M. Raza, E. Prieto-Araujo and O. Gomis-Bellmunt, "Small-Signal Stability Analysis of Offshore AC Network Having Multiple VSC-HVDC Systems," in *IEEE Transactions on Power Delivery*, vol. 33, no. 2, pp. 830-839, April 2018.
[9]. L. Xu, L. Fan and Z. Miao, "DC Impedance-Model-Based Resonance Analysis of a VSC–HVDC System," in IEEE Transactions on Power Delivery, vol. 30, no. 3, pp. 1221-1230, June 2015.
[10]. Belkhayat M, "Stability criteria for AC power systems with regulated loads," Ph.D. dissertation, Purdue University, USA, 1997.
[11]. B. Wen, D. Boroyevich, R. Burgos, P. Mattavelli and Z. Shen, "Small-Signal Stability Analysis of Three-Phase AC Systems in the Presence of Constant Power Loads Based on Measured d-q Frame Impedances," in IEEE Transactions on Power Electronics, vol. 30, no. 10, pp. 5952-5963, Oct. 2015.
[12]. L. Harnefors, M. Bongiorno and S. Lundberg, "Input-Admittance Calculation and Shaping for Controlled Voltage-Source Converters," in *IEEE Transactions on Industrial Electronics*, vol. 54, no. 6, pp. 3323-3334, Dec. 2007.
[13]. A. Rygg, M. Molinas, C. Zhang and X. Cai, "A Modified Sequence-Domain Impedance Definition and Its Equivalence to the dq-Domain Impedance Definition for the Stability Analysis of AC Power Electronic Systems," in IEEE Journal of Emerging and Selected Topics in Power Electronics, vol. 4, no. 4, pp. 1383-1396, Dec. 2016.
[14]. C. Zhang, X. Cai, A. Rygg and M. Molinas, "Sequence Domain SISO Equivalent Models of a Grid-Tied Voltage Source Converter System for Small-Signal Stability Analysis," in IEEE Transactions on Energy Conversion, vol. 33, no. 2, pp. 741-749, June 2018.
[15]. S. Shah and L. Parsa, "Impedance Modeling of Three-Phase Voltage Source Converters in DQ, Sequence, and Phasor Domains," in IEEE Transactions on Energy Conversion, vol. 32, no. 3, pp. 1139-1150, Sept. 2017.
[16]. M. Cespedes and J. Sun, "Impedance Modeling and Analysis of Grid-Connected Voltage-Source Converters," in IEEE Transactions on Power Electronics, vol. 29, no. 3, pp. 1254-1261, March 2014.
[17]. M. K. Bakhshizadeh, X. Wang, F. Blaabjerg, J. Hjerrild, L. Kocewiak, C. L. Bak, and B. Hesselbæk, "Couplings in Phase Domain Impedance Modeling of Grid-Connected Converters," IEEE Trans. Power Electron, vol. 31, no. 10, pp. 6792–6796, 2016.
[18]. X. Wang and F. Blaabjerg, "Harmonic Stability in Power Electronic Based Power Systems: Concept, Modeling, and Analysis," in *IEEE Transactions on Smart Grid*. doi: 10.1109/TSG.2018.2812712 *(online)*.
[19]. C. Desoer and Yung-Terng Wang, "On the generalized nyquist stability criterion," in IEEE Transactions on Automatic Control, vol. 25, no. 2, pp. 187-196, April 1980
[20]. J. Sun. "Impedance-Based Stability Criterion for Grid-Connected Inverters," IEEE Trans. Power Electron, vol.26, no. 11, pp. 3075–3078, 2011.
[21]. H. Liu and X. Xie, "Impedance Network Modeling and Quantitative Stability Analysis of Sub-/Super-Synchronous Oscillations for Large-Scale Wind Power Systems," in *IEEE Access*, vol. 6, pp. 34431-34438, 2018.

[22]. E. Ebrahimzadeh, F. Blaabjerg, X. Wang and C. L. Bak, "Harmonic Stability and Resonance Analysis in Large PMSG-Based Wind Power Plants," in IEEE Transactions on Sustainable Energy, vol. 9, no. 1, pp. 12-23, Jan. 2018.
[23]. C. Zhang, X. Cai, Z. Li, A. Rygg and M. Molinas, "Properties and physical interpretation of the dynamic interactions between voltage source converters and grid: electrical oscillation and its stability control," in IET Power Electronics, vol. 10, no. 8, pp. 894-902, 30 6 2017.
[24]. S. Ma, H. Geng, L. Liu, G. Yang and B. C. Pal, "Grid-Synchronization Stability Improvement of Large Scale Wind Farm During Severe Grid Fault," in *IEEE Transactions on Power Systems*, vol. 33, no. 1, pp. 216-226, Jan. 2018.
[25]. L. Harnefors, "Modeling of Three-Phase Dynamic Systems Using Complex Transfer Functions and Transfer Matrices," IEEE Trans. Ind. Electron, vol. 54, no. 4, pp. 2239–2248, 2007.
[26]. G. C. Paap, "Symmetrical components in the time domain and their application to power network calculations," in *IEEE Transactions on Power Systems*, vol. 15, no. 2, pp. 522-528, May 2000.
[27]. C. Zhang, X. Cai, M. Molinas and A. Rygg, "On the Impedance Modeling and Equivalence of AC/DC Side Stability Analysis of a Grid-tied Type-IV Wind Turbine System," in IEEE Transactions on Energy Conversion. doi: 10.1109/TEC.2018.2866639 (online).
[28]. M. Amin, M. Molinas, J. Lyu and X. Cai, "Impact of Power Flow Direction on the Stability of VSC-HVDC Seen From the Impedance Nyquist Plot," in IEEE Transactions on Power Electronics, vol. 32, no. 10, pp. 8204-8217, Oct. 2017.
[29]. B. Wen, D. Boroyevich, R. Burgos, P. Mattavelli and Z. Shen, "Inverse Nyquist Stability Criterion for Grid-Tied Inverters," in *IEEE Transactions on Power Electronics*, vol. 32, no. 2, pp. 1548-1556, Feb. 2017.